\documentstyle[psfig,aaspp4,12pt]{article}
 
\slugcomment{to appear in the Astrophysical Journal} 
 
\begin{document}
 
\title{Variability of sub-mJy radio sources}
   
\author{C. L. Carilli}
\affil{National Radio Astronomy Observatory, P.O. Box O, Socorro, NM,
87801, USA \\
ccarilli@nrao.edu}
\author{R.J. Ivison}
\affil{Astronomy Technology Centre, Royal Observatory, Blackford
Hill, Edinburgh EH9 3HJ, UK}
\author{D.A. Frail}
\affil{National Radio Astronomy Observatory, P.O. Box O, Socorro, NM,
87801, USA \\}

\begin{abstract}

We present 1.4 GHz VLA observations of the variability of radio
sources in the Lockman Hole region at the level of $\ge 100\mu$Jy on
timescales of 17 months and 19 days.  These data indicate that the
areal density of highly variable sources at this level is $<
5\times10^{-3}$ arcmin$^{-2}$.  We set an upper limit of $2\%$ to the
fraction of 50 to 100$\mu$Jy sources that are highly variable ($\rm
\Delta S \ge 50\%$).  These results imply a lower limit to the beaming
angle for GRBs of 1$^o$, and give a lower limit of 200 arcmin$^2$ to
the area that can be safely searched for GRB radio afterglows before
confusion might become an issue.

\end{abstract}


\section {Introduction}

Synoptic surveys for Gamma-ray bursts (GRBs), and subsequent
ground-based observational follow-up at radio through optical
wavelengths, has highlighted the importance of transient celestial
phenomena (Masetti 2001).  The new parameter space of the transient
cosmos has been emphasized in the design of future telescopes, such as
the optical Large Synoptic Survey Telescope (Tyson \& Angel 2001), and
the radio Square Kilometer Array (van Haarlem 1999).

While it is well documented that flat spectrum radio sources can be
variable (Aller et al. 1985), the areal density of such sources has
not been well quantified through multi-epoch, wide field blind
surveys. At high flux density levels ($> 10$ mJy at 1.4 GHz), one
can make a rough estimate of the areal density of variable radio
sources by simply assuming that all flat spectrum sources are
variable. For instance, the areal density of all sources $\ge 10$mJy
is $\sim 2\times10^{-3}$ arcmin$^{-2}$, and the fraction of flat
spectrum sources is about 10$\%$, implying an areal density of
variable radio sources of $\sim 2\times10^{-4}$ arcmin$^{-2}$
(Gruppioni et al. 1999; White et al. 1997; Windhorst et al. 1985;
Hopkins et al. 2000, 2002).  This number is consistent with the (null)
results of Frail et al. (1994) in their search for highly variable
mJy-level sources associated with GRBs.

Source populations at these high flux densities are dominated by AGN.
Below about 1 mJy the slope of the source counts flattens, and star
forming galaxies are thought to dominate the faint source population
(Windhorst et al. 1985; Georgakakis et al. 1999; Hopkins et al. 2000,
2002). Hence, when considering the areal density of variable sub-mJy
radio sources, one cannot simply extrapolate the results from high
flux density source samples to low flux densities.

Knowledge of the areal density of variable sub-mJy radio sources is
critical for setting the back-ground, or `confusion', level for studies
of faint variable source populations, such as GRBs (Frail et
al. 1997).  A recent comparison of the NVSS and FIRST surveys by
Levinson et al. (2002) sets a conservative upper limit of
$3\times10^{-6}$ arcmin$^{-2}$ to the areal density of `orphan' GRB
radio afterglows (i.e. GRBs for which the $\gamma$-ray emission is not
beamed toward us) with S$_{1.4} \ge 6$ mJy.  They also argue that the
areal density of radio supernovae will be considerably smaller.  While
the area of the sky covered by Levinson et al. (2002) was much larger
that the study presented herein, their flux density limit was higher
than any GRB radio afterglow yet recorded.  In this paper we present a
smaller area study, but we consider variable sources at flux density
levels ($\sim 0.1$mJy) applicable to typical GRB radio afterglows.

In general, variability of radio sources at the sub-mJy level is an
essentially unexplored part of parameter space -- a part of parameter
space which may fundamentally drive the design of future radio
telescopes, such as the SKA (Carilli et al. 2002).  Herein we present
the first study to delve into this part of parameter space, by
exploring systematically the variability of the sub-mJy radio source
population at 1.4 GHz. We examine variability on timescales of 17
months and 19 days.  Note that the LSST will probe similar variability
timescales in the optical, with sampling on weekly to yearly
timescales. 

\section{Observations}

Observations were made using the using the VLA at 1.4 GHz in the B
configuration (maximum baseline = 10km). The region observed is in the
Lockman Hole centered at: (J2000) 10$^h$ 52$^m$ 56.00$^s$, 57$^d$
29$'$ 06.0$''$.  Table 1 summarizes the observations. Column 1 gives
the observing date, column 2 gives the observed hour angle range, and
column 3 gives the rms in the final image.  These observations are
part of a larger multiwavelength program to study the evolution of
dusty star forming galaxies (Bertoldi et al. in prep).

Standard wide field imaging techniques were employed in order to
generate an unaberrated image of the full primary beam of the VLA
(FWHM = 32$'$). The absolute flux density scale was set using
3C286. We then generated a CLEAN component model of the field using
self-calibrated data taken on Sept. 5, 2002.  The data from all days
were then self-calibrated in amplitude and phase without gain
renormalization using this model. This process should ensure that all
the data are on the same flux scale.  Images made before and after
this process showed that the absolute flux scale changed by at most
1$\%$.  We also checked to see if variable sources could be removed
(or added) due to using a single model to self-calibrate data from all
the different days.  Components at the 0.1 to 0.2 mJy level were added
to the self-calibration model at random positions in the field, and
the self-calibration process was repeated.  In no case was a new
source generated. This gives us confidence that the self-calibration
process is robust to small perturbations in the model, i.e. that the
problem is over-constrained and that the input self-calibration model
is dominated by non-varying sources.  

Images for each day were generated using the wide field imaging
capabilities in the AIPS task IMAGR (Perley 1999).  To remove problems
with `beam squint' (slightly different pointing centers for right and
left circular polarizations) the right and left polarizations were
imaged separately.  The images were then summed, weighted by the rms
on each image.  The final image using data from all the observing days
is shown in Figure 1. The rms noise on this image is 7$\mu$Jy
beam$^{-1}$ and the restoring CLEAN beam is circular with FWHM =
4.5$''$.

\section{Analysis}

We searched for source variations over 17 months by comparing images
made in April 2001 with those made in August/September 2002, and also
on timescales of 19 days by comparing images made on August 17 and
September 5, 2002. We searched for variable sources out to the 10$\%$
point of the primary beam, and we limited the analysis to sources with
variations $>2\%$.  Images from all epochs were convolved to 6$''$
resolution to mitigate differences that might occur due to the
non-linear process of deconvolution. The rms on the images used for
variability analysis was 12.5 $\mu$Jy for the 17 month comparison, and
17$\mu$Jy for the 19 day comparison. 

The images from different epochs were both subtracted and averaged. A
fractional variability image was then generated by dividing these two
images, blanking at the 5$\sigma$ level (in absolute value).  For
the 17 months variability analysis $\sigma$ was 17$\mu$Jy on the
differenced image, and 9$\mu$Jy on the averaged image. The
corresponding numbers for the 19 day analysis were 26$\mu$Jy and
15$\mu$Jy, respectively. 

The difference image for the 17 month analysis is shown in Figure 2.
Some artifacts are seen around the brighter extended sources arising
from differences in deconvolution and residual calibration errors.
Beyond these artifacts, the difference images are remarkably free of
sources. This result gives us confidence that the imaging process
(self-calibration and deconvolution) does not generate spurious
sources at the $\ge 5\sigma$ level, and tells us right away that the
radio sky is not highly variable at the 100$\mu$Jy level.

Variable sources were identified in the divided image. Five variable
sources were found at the $5\sigma$ level in the 17 month comparison,
while four sources were found in the 19 day analysis.  We then
returned to the original images to find the flux densities of the
sources at each epoch.  Table 2 lists source positions (columns 1 and
2), flux densities at the two epochs (columns 3 and 4), and the
distance from the phase center (column 5) for the 17 month analysis,
and Table 3 lists the corresponding values for the 19 day analysis.
Note that the source J1051+5734 was seen to vary on both
timescales. In fact, this source varied between September 5 and
September 9 from 1.53 mJy to 0.71 mJy. The September 2002 value listed
in Table 2 is the weighted average of these two measurements.

We next consider the sensitivity of our observations to variable
sources at some absolute level, $\Delta$S. The analysis is complicated
by the roughly Gaussian roll-off of the primary beam of the VLA, with
FWHM $\sim 32'$.  Identification of variable sources was done using
the non-primary beam corrected maps in order to have uniform noise
across the field.  Of course, the final flux densities and noise
levels for the variable sources were corrected for the primary beam
attenuation.

Given a $\Delta$S, one can calculate the maximum primary beam
correction, f, for which a change in flux density $\Delta$S could be
detected at the 5$\sigma$ level, where $\sigma$ is the noise at the
field center on the difference image: f = 5$\sigma$/$\Delta$S. The
value of f then sets the distance from the pointing center, R, 
to which such variation could have been detected, given the 
primary beam shape of the VLA.

Values for $\Delta$S, f, and R are listed in columns 1, 2, and 3,
respectively in Table 4. Column 4 lists the number of variable sources
over 17 months that meet these criteria. 
Column 5 lists the number of sources within the specified
radius with flux density, S $\ge$ $\Delta$S/2.  This flux
density sets the limit for 100$\%$ variability, e.g. a 
source could be 100$\mu$Jy on day 1, and 0 $\mu$Jy on day 2,
leading to a value on the difference image of 
100 $\mu$Jy and a value on the average image of 50$\mu$Jy. 
Column 6 lists the areal density of sources with S$\ge$$\Delta$S/2
on our Lockman hole image, while column 7 lists the corresponding
values from the recent study of Fomalont et al. (2002) comprised
of a number of different fields. 

We have investigated the source sizes using the B array observations
at 4.5$''$ resolution.  Two of the sources are partially resolved.
For J1051+5708 Gaussian fitting to the profile yields a peak surface
brightness, I$_\nu = 8.92\pm0.05$ mJy beam$^{-1}$, a total flux density,
S$_\nu = 10.00\pm0.09$ mJy, and a (deconvolved) size of
$1.9''\times1.1''$, with major axis position angle (PA) = $37^o$. The
corresponding numbers for J1055+5718 are: I$_\nu = 8.82\pm0.06$
mJy beam$^{-1}$, S$_\nu = 12.35\pm0.13$ mJy, and 
$3.4''\times 2.2''$ at PA $= 81^o$.  The rest of the sources are
unresolved, with upper limits between 1$''$ and 2$''$ depending on
signal-to-noise. 

The nature of this analysis is such that we are not 
sensitive to very rapid variations, e.g. timescales of minutes
or less. For instance, a 50 mJy flare of 1 min duration would
average down to about 100$\mu$Jy over 7 hours. While our
nominal sensitivity would be adequate to detect such an
event, the existence of such a transient source in the visibility 
data would lead to errors in the self-calibration and
imaging process which would manifest themselves clearly on
the images. Of course, it is possible that such a 
bright, short timescale event was mis-identified as interference
in the data editing process, and removed. 

\section{Discussion}

For the analysis on 17 month timescales we could have detected sources
with $\Delta$S $\ge 100\mu$Jy out to a radius of 7.8$'$, but none were
detected.  This sets an upper limit to the areal density of such
sources of $5\times10^{-3}$ arcmin$^{-2}$.  Another interesting point
is that within this radius there are 46 sources between 50 and 100
$\mu$Jy on the averaged image. Hence, we set an upper limit of about
2$\%$ to the fraction of sources in this flux density regime that are
highly ($\ge 50\%$) variable.

These results are grossly consistent with the idea that below 1 mJy at
1.4 GHz the radio source population is dominated by star forming
galaxies, as compared to AGN which dominate at high flux densities
(Hopkins 2000, 2002). The exact distribution of AGN vs. starbursts
vs. other source types as a function of radio flux density is not
fully determined at this time, and is an area of active current
research (Hopkins et al. 2002; Georgakakis et al. 1999; Richards 2001;
Fomalont et al. 2002).  As a rough guide we consider the models of
Hopkins et al. (2000). They suggests that at $\ge$ 10 mJy the source
population is 90$\%$ steep spectrum radio sources, 10$\%$ flat
spectrum radio sources, and $\le 1\%$ star forming galaxies.  Again,
at high flux density levels the flat spectrum sources correspond to
the variable radio source population.  At 100 $\mu$Jy the models
suggests that the proportions change to roughly 80$\%$ star forming
galaxies, 15$\%$ steep spectrum AGN, and 5$\%$ flat spectrum AGN.

These models have not considered truly transient source populations,
such as GRB radio afterglows, which have timescales of 10's of days.
The statistics of radio afterglows of GRBs are such that in an area of
7.8$'$ radius one would expect to see $\sim 0.005$ sources above 0.1
mJy at 1.4 GHz at any given time, assuming GRBs are highly beamed (see
below). Hence, the fact that we did not see such a source is not
surprising.  More importantly, the results presented herein allow us
to set a limit on the variable source confusion level for GRB radio
afterglow searches, again at a flux density level relevant to the
observed population. For example, radio searches are needed for
localizing an important subclass of afterglows known as `dark GRBs'
(e.g., Djorgovski et al.~2001), for which optical emission from the
GRB is not detected. It has been suggested that the absence of optical
emission from these sources is the result of either dust obscuration,
or the Gunn-Peterson effect, i.e. Ly$\alpha$ absorption by the neutral
intergalactic medium. This latter effect would place the sources at $z
> 6.3$ (Fan et al. 2003).  An upper limit to the areal density of
$5\times10^{-3}$ arcmin$^{-2}$ for variable sources at the 100$\mu$Jy
level implies a lower limit of 200 arcmin$^{2}$ to the area that can
be safely searched at 1.4 GHz before one such source is expected to be
detected by chance. For comparison, the typical GRB error circle is
$\sim$30 arcmin$^{2}$, but larger error circles are not uncommon.

This result can also be used to derive a rough limit on the mean
beaming angle for GRBs (Perna \& Loeb 1998). From a sample of 25 radio
afterglows (Frail et al. in prep) we estimate that 10-25\% will be
visible above the 100 $\mu$Jy level at 1.4 GHz, with an average
lifetime of one month. The GRB event rate is approximately 600 per
year (Fishman \& Meegan 1995) so we expect to find only
$8.4\times10^{-8}$ arcmin$^{-2}$.  However, if GRBs are highly beamed,
as recent studies seem to suggest (Frail et al.~2001), then our upper
limit to areal density of $5\times10^{-3}$ arcmin$^{-2}$ implies a
beaming factor $f_b < 6\times10^4$, or a mean jet opening angle
$\theta_j > 1^\circ$.  This value is not very constraining compared to
existing limits, but to our knowledge this is the first time a survey
for variability has been done at the appropriate flux level for radio
afterglows. More stringent limits will require sensitive, larger area
surveys with existing or planned instruments (Totani \& Panaitescu
2002).

A final point we consider is cosmic variance.  It is possible that the
Lockman hole region we have sampled was just statistically-poor in
transient sources. A rough indication of the effect of cosmic variance
comes from the overall source counts (columns 6,7 in Table 4). The
source counts we derive agree to within 20$\%$ with those found by
Fomalont et al. (2002) in other areas of the sky. In general, it has
been found that for 30$'$ fields-of-view the maximum field-to-field
scatter in the sub-mJy source counts is about a factor two (Fomalont
et al. 2002). We consider this an upper limit to the effect cosmic
variance has on the areal density of variable source presented herein.

\vskip 0.2truein

The National Radio Astronomy Observatory (NRAO) is operated by
Associated Universities, Inc. under a cooperative agreement with the
National Science Foundation. We thank F. Owen, E. Fomalont, R. Becker,
and J. Condon for useful comments concerning this work, and
the referee for careful and insightful criticism.

\vfill\eject

\clearpage

\begin{table}[htb]
\caption{Observations} 
\vskip 0.2in
\begin{tabular}{lll}
\hline
\hline
Date & Hour Angle Range & RMS \\  
~ & ~ & $\mu$Jy \\
\hline
April 22, 2001  & -4.2 to +1.0 & 15 \\
April 27, 2001  & -4.2 to +1.3 & 15 \\
August 17, 2002  &  -2.2 to +3.4 & 17 \\
September 5, 2002 & -3.2 to +3.7 & 16 \\
September 9, 2002 & -2.8 to +3.1 & 18 \\
\hline
\end{tabular}
\end{table}

\clearpage\newpage

\begin{table}[htb]
\caption{Variable sources: April 2001 to Sept 2002} 
\vskip 0.2in
\begin{tabular}{lllll}
\hline
\hline
RA & Dec & I$_{\rm April 2001}$ & I$_{\rm Sept 2002}$ & Radius \\ 
J2000 & J2000 & mJy beam$^{-1}$ & mJy beam$^{-1}$ & arcmin  \\
\hline
10 50 39.53 & 57 23 36.5 & $5.70\pm0.036$ & $4.48\pm0.036$ & 19.2 \\
10 51 22.06 & 57 08 54.8 & $9.47\pm0.07$ &  $9.98\pm0.07$ &  23.8 \\ 
10 51 42.03 & 57 34 47.7 & $0.80\pm0.018$ & $1.18\pm0.018$ & 11.4 \\
10 54 00.48 & 57 33 21.2 & $2.79\pm0.016$ & $2.66\pm0.016$ & 9.7 \\
10 55 48.52 & 57 18 27.5 & $10.96\pm0.10$ & $10.08\pm0.10$ & 25.6 \\
\hline
\end{tabular}
\end{table}

\begin{table}[htb]
\caption{Variable sources: August 17 to Sept 5, 2002} 
\vskip 0.2in
\begin{tabular}{lllll}
\hline
\hline
RA & Dec & I$_{\rm Aug~17}$ & I$_{\rm Sept~5}$ & Radius \\ 
J2000 & J2000 & mJy beam$^{-1}$ & mJy beam$^{-1}$ & arcmin  \\
\hline
10 51 38.08 & 57 49 56.5 & $1.97\pm0.099$ & $3.02\pm0.099$ & 23.3 \\
10 51 42.03 & 57 34 47.7 & $1.22\pm0.027$ & $1.53\pm0.027$ & 11.4  \\
10 52 06.44 & 57 41 09.7 & $9.24\pm0.031$ & $9.58\pm0.031$ & 13.8 \\
10 52 25.39 & 57 55 05.6 & $23.4\pm0.18$  & $25.4\pm0.18$ & 26.3 \\
\hline
\end{tabular}
\end{table}

\vfill \eject

\begin{table}[htb]
\caption{Search radius for variable sources at 1.4 GHz} 
\vskip 0.2in
\begin{tabular}{ccccccc}
\hline
\hline
$\Delta$S & f$^a$ & R$^b$ & N$_{\rm var}^c$  & 
N$_{\rm srcs}^d$ & N$_{\rm srcs}$/area$^e$ & N$_{\rm
srcs}$/area$^f$  \\ 
mJy & ~ & arcmin & ~ & $\rm S\ge\Delta S/2$ 
& $\rm S\ge\Delta S/2$ arcmin$^{-2}$ &  $\rm S\ge\Delta S/2$
arcmin$^{-2}$ \\
\hline
0.13 & 0.65 & 12.7 & 1 & 219 & 0.43 & 0.53 \\
0.3 & 0.28 & 20.9 & 1  & 222 & 0.16 & 0.21 \\
0.50 & 0.17 & 23.9 & 1 & 187 & 0.10 & 0.12 \\
0.85 & 0.1 & 27.0 & 1 & 139 & 0.061 & 0.067 \\
\hline
\end{tabular}
\end{table}
~$^a$f = 5$\sigma/\Delta$S, where $\sigma = 17\mu$Jy = rms on
difference maps at the field center. 

~$^b$R = distance from pointing center out to which $\Delta$S can be
detected to $\ge 5\sigma$.

~$^c$N$_{\rm var}$ = number of variable sources $\ge \Delta$S
within given radius over 17 months.

~$^d$N$_{\rm srcs}$ = number of sources with flux density,
$\rm S \ge \Delta S/2$ within R. This S sets the 100$\%$ variability
detection limit.  

~$^e$N$_{\rm srcs}$/area = number of sources with $\rm S \ge \Delta
S/2$ per arcmin$^{2}$ from our Lockman Hole field. 

~$^f$N$_{\rm srcs}$/area = number of sources with $\rm S \ge \Delta
S/2$ per arcmin$^{2}$ from the study of Fomalont et al. (2002):
N($\rm \ge S_{mJy}) = 0.026\times S^{-1.1}~ arcmin^{-2}$.

\vfil\eject

\vfill\eject

\centerline{\bf Figure Captions}

F{\scriptsize IG}. {\bf 1}.--- Radio image of the Lockman Hole
at 1.4 GHz with a resolution of 4.5$''$. The rms noise on the
image is 7$\mu$Jy. The grayscale range is from -0.15 mJy to 0.25 mJy.

F{\scriptsize IG}. {\bf 2}.--- The difference image between 
observations in April 2001 and Aug/Sept 2002 at 6$''$ resolution. 
The grayscale range is from -0.2 mJy to 0.2 mJy.



\begin{figure}
\psfig{figure=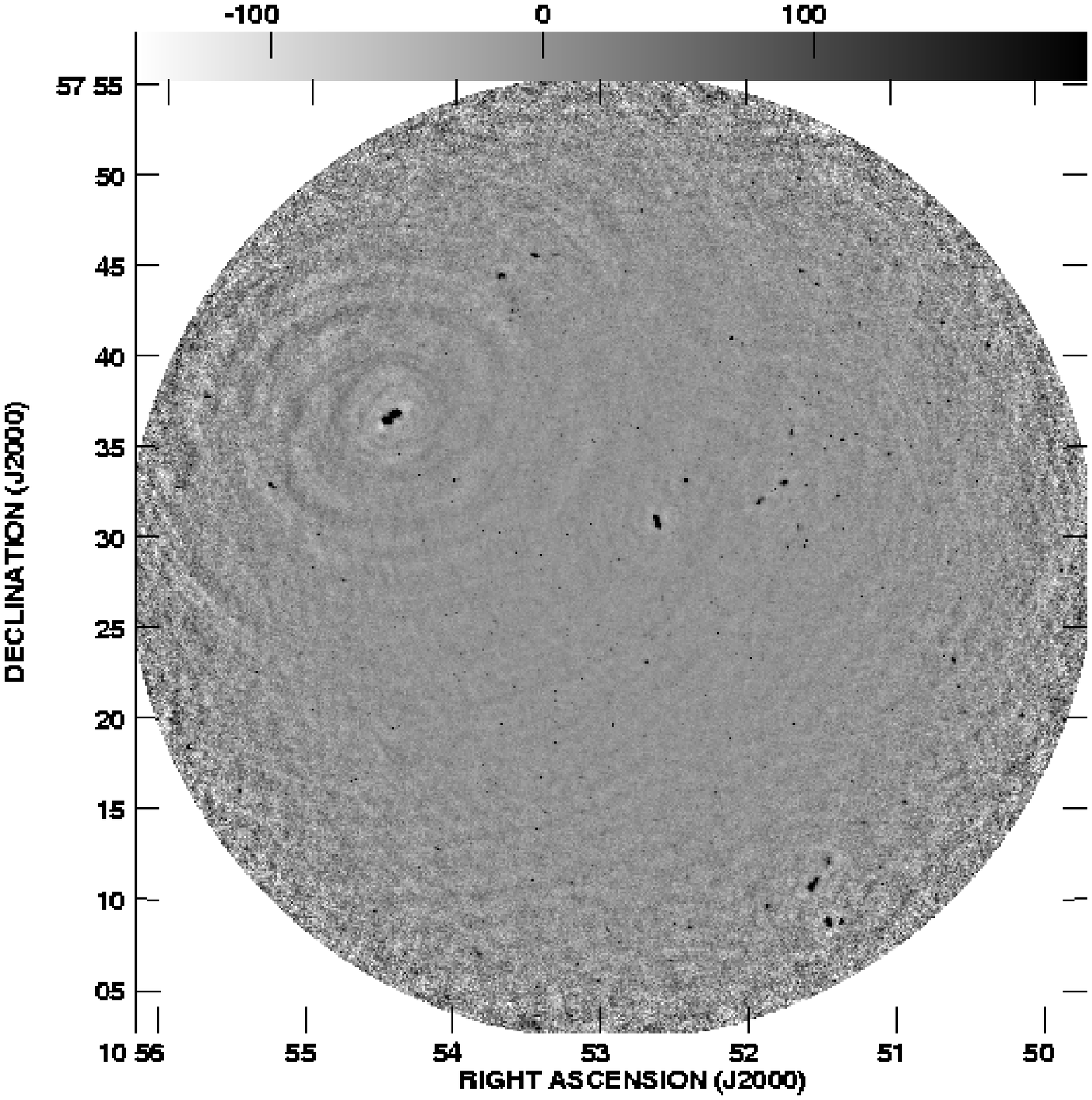,width=7in}
\caption{}
\end{figure}

\clearpage
\newpage

\begin{figure}
\psfig{figure=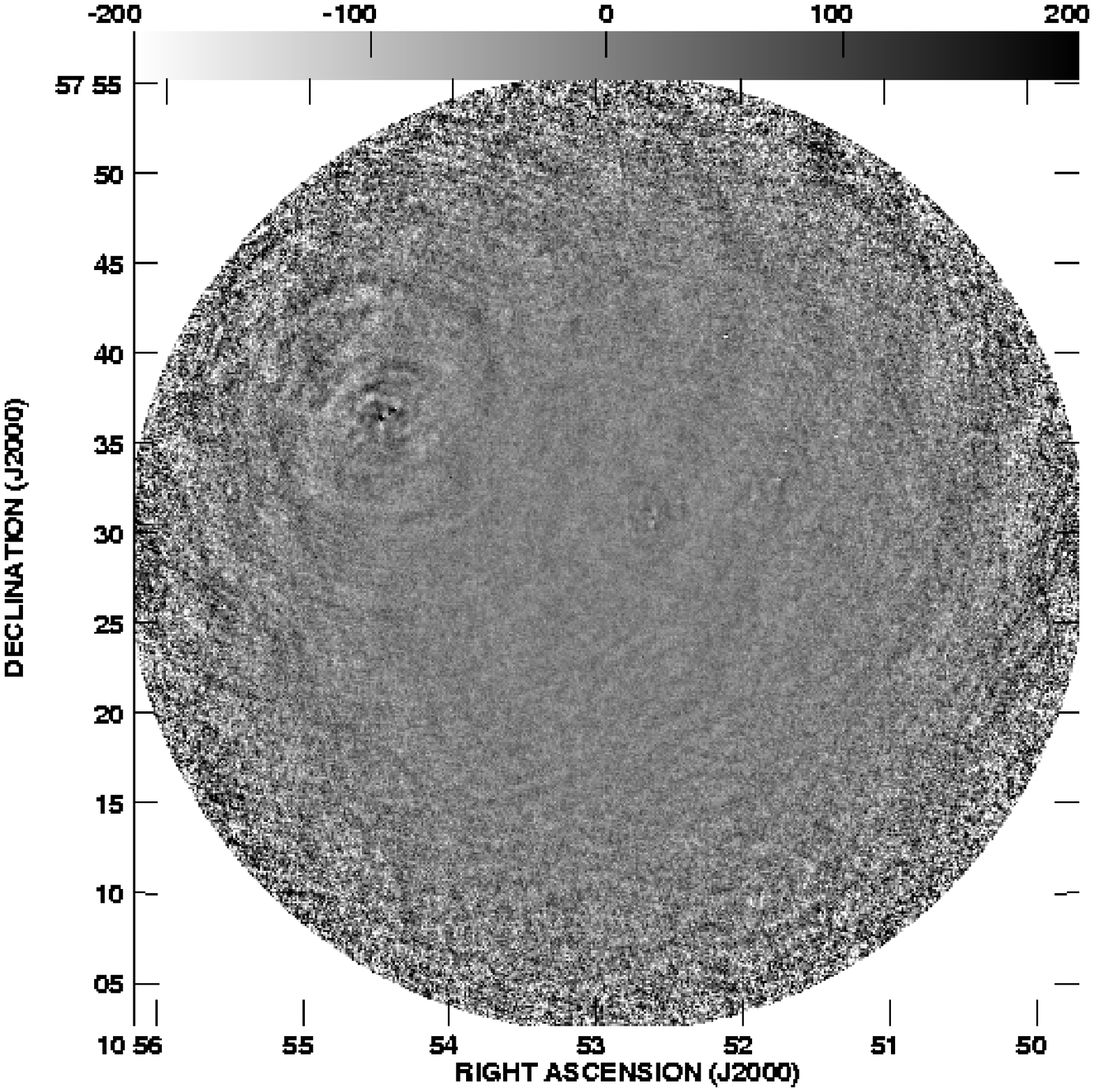,width=7in}
\caption{}
\end{figure}

\end{document}